\documentclass[final,3p,times]{elsarticle}
\usepackage{amsmath}
\usepackage{lineno,hyperref}

\modulolinenumbers[5]

\biboptions{sort&compress}
\journal{Annals of Physics}

\hypersetup{colorlinks=true,linkcolor=blue,citecolor=blue,filecolor=blue,urlcolor=blue,breaklinks=true}

\begin{document}

\begin{frontmatter}

\title{Pseudospin and spin symmetries in 1+1 dimensions:\\ The case of the Coulomb potential}

\author[mymainaddress]{Luis B. Castro\corref{mycorrespondingauthor}}
\cortext[mycorrespondingauthor]{Corresponding author}
\ead{lrb.castro@ufma.br}

\author[mysecondaryaddress]{Antonio S. de Castro}
\ead{castro@pq.cnpq.br}

\author[mythirdaddress]{Pedro Alberto}
\ead{pedro.alberto@uc.pt}

\address[mymainaddress]{Departamento de F\'{\i}sica, Universidade Federal do Maranh\~{a}o, Campus Universit\'{a}rio do Bacanga,\\ 65080-805, S\~{a}o Lu\'{\i}s, MA, Brazil.}
\address[mysecondaryaddress]{Departamento de F\'{\i}sica e Qu\'{\i}mica, Universidade Estadual Paulista, Campus de Guaratin\-gue\-t\'{a},\\ 12516-410, Guaratinguet\'{a}, SP, Brazil.}
\address[mythirdaddress]{Centro de F\'{\i}sica Computacional, Physics Department of the University of Coimbra,\\
P-3004-516, Coimbra, Portugal}

\begin{abstract}
The problem of fermions in 1+1 dimensions in the presence of a pseudoscalar
Coulomb potential plus a mixing of vector and scalar Coulomb potentials
which have equal or opposite signs is investigated. We explore all the
possible signs of the potentials and discuss their bound-state solutions for
fermions and antifermions. We show the relation between spin and pseudospin
symmetries by means of charge-conjugation and $\gamma^{5}$ chiral
transformations.
The cases of pure pseudoscalar and mixed vector-scalar potentials, already analyzed in
previous works,
are obtained as particular cases. The results presented can be extended to 3+1
dimensions.
\end{abstract}

\begin{keyword}
Pseudospin and spin symmetry \sep Charge conjugation \sep Chiral transformation \sep Coulomb potential
\PACS 21.10.Hw \sep 03.65.Ge \sep 03.65.Pm
\end{keyword}

\end{frontmatter}


\section{Introduction}

The potential generated by a point charge, the Coulomb potential, depends on the
dimensionality of space-time (see, e.g.
\cite{Efthimiou:2000}). The (1+1)-dimensional Coulomb potential is linear and so it
provides a constant electric field
always pointing to, or from, the point charge. This problem is related to
the confinement of fermions in the Schwinger and in the massive Schwinger
models \cite{AP93:267:1975,AP101:239:1976} as well as in the Thirring-Schwinger model \cite{HPA49:889:1976}.%
Considered as the time component of a Lorentz vector and due to the tunneling effect
(Klein\'{}s paradox), there are no bound states for this kind of potential regardless of the
strength of the potential \cite{CJP63:1029:1985,AJP56:312:1988}. The linear potential,
considered as a Lorentz scalar, is also related to the quarkonium model in
1+1 dimensions \cite{NPB75:461:1974,PRD9:3501:1974}. Although it was incorrectly
concluded that even in this case there is just one bound state \cite{AJP69:817:2001},
later on the proper solutions for this last problem were found \cite{AJP70:450:2002,AJP70:451:2002,AJP70:522:2002}. However, it is well known from the quarkonium phenomenology in the
real 3+1 dimensional world that the best fit for meson spectroscopy is found
for a convenient mixture of vector and scalar potentials put by hand in the
equations (see, e.g., \cite{PR200:127:1991}). The same can be said about the treatment
of the nuclear phenomena describing the influence of the nuclear medium on
the nucleons \cite{WALECKA1986}. The mixed vector-scalar potential has
also been analyzed with the Dirac equation in 1+1 dimensions for a linear
potential \cite{PLA305:100:2002} as well as for a general potential which goes to
infinity as $|x|\rightarrow \infty $ \cite{AJP71:950:2003}. In both of those last
references it has been concluded that there is confinement if the scalar
coupling is of sufficient intensity compared to the vector coupling. The
problem has also been analyzed for pseudoscalar couplings \cite{PLA318:40:2003}.

The case in which the mean field is composed of a vector ($V_{t}$) and a
scalar ($V_{s}$) potentials, with $V_{t}=-V_{s}$, is particularly relevant
in nuclear physics, because it is usually pointed out as a necessary
condition for occurrence of pseudospin symmetry in nuclei \cite%
{PRL78:436:1997,PR315:231:1999,PRC58:R628:1998,PRC62:054309:2000,PRL86:5015:2001,PRC65:034307:2002,
PR414:165:2005,PRL109:072501:2012,PRC87:014334:2013,PRL112:062502:2014,Liang:2014}.
Furthermore, with an appropriate choice of
signs, potentials fulfilling the relations $V_{s}=\pm V_{t}$ are able to
bind either fermions or antifermions \cite{AP316:414:2005,PRC81:064324:2010,PRA86:032122:2012,PRC86:052201:2012,JPA46:085305:2013}.

Closely related to this is the fact that, in the nucleus, the
charge-conjugation transformation relates the spin symmetry of the negative
bound-state solutions (antinucleons) to the pseudospin symmetry of the
positive bound-state solutions (nucleons) \cite{PRC81:064324:2010,PRL91:262501:2003}. Therefore, we believe
that this connection between spin and pseudospin symmetry obtained by charge
conjugation deserves to be more explored.

The main motivation of this article is illustrate the relation
between spin and pseudospin symmetries using charge-conjugation and
chiral transformations in the Dirac equation in 1+1 dimensions, which was already uncovered in
ref.~\cite{PRC73:054309:2006}, for the problem of mixed
scalar-vector-pseudoscalar Coulomb potential in 1+1 dimensions. We thus take
advantage of the simplicity of the lowest dimensionality of the space-time
as much as was done for the harmonic oscillator potential \cite{PRC73:054309:2006} and P%
\"{o}schl-Teller potential \cite{IJMPE16:3002:2007}. This approach is equivalent to
consider fermions in 3+1 dimensions that are restricted to move in one
direction \cite{STRANGE1998}. We explore the spectra when it is possible to obtain
analytical solutions, i.e., in the particular cases when $\Delta=V_{t}-V_{s}=0$ or $%
\Sigma=V_{t}+V_{s}=0$. We explore all the possible signs of the potentials, thus paying
attention to bound states of fermions and antifermions as well. We compare
both cases $\Delta=0$ and $\Sigma=0$ to establish the charge-conjugation
connection discussed above in the presence of the pseudoscalar term. We also
discuss the connection between spin and
pseudospin symmetries by means of the chiral transformation.

\section{The Dirac equation in 1+1 dimensions}

The 1+1 dimensional time-independent Dirac equation for a fermion of rest
mass $m$ under the action of vector ($V_{t}$), scalar ($V_{s}$) and
pseudoscalar ($V_{p}$) potentials can be written, in terms of the
combinations $\Sigma =V_{t}+V_{s}$ and $\Delta =V_{t}-V_{s}$, as
\begin{equation}
H\psi =E\psi ,\quad H=c\sigma _{1}p+\sigma _{3}mc^{2}+\frac{I+\sigma _{3}}{2}%
\Sigma +\frac{I-\sigma _{3}}{2}\Delta +\sigma _{2}V_{p}\,,  \label{eq1}
\end{equation}

\noindent where $E$ is the energy of the fermion, $c$ is the velocity of
light and $p$ is the momentum operator. The matrices $\sigma _{1}$, $\sigma
_{2}$ and $\sigma_{3}$ denote the Pauli matrices, and $I$ denotes the $%
2\times 2$ unit matrix. The positive definite function $|\psi |^{2}=\psi
^{\dagger }\psi $, satisfying a continuity equation, is interpreted as a
position probability density and its norm is a constant of motion. This
interpretation is completely satisfactory for single-particle states \cite%
{THALLER1992}.

\subsection{Parity, charge conjugation and chiral transformation}

The Dirac equation is covariant under $x\rightarrow -x$ if \ $V_{p}$ changes
sign whereas $V_{s}$ and $V_{t}$ remain the same. This is because the parity
operator $P=\exp \left( i\varepsilon \right) P_{0}\sigma _{3}$, where $%
\varepsilon $ is a constant phase and $P_{0}$ changes $x$ into $-x$, changes
the sign of $\sigma _{1}$ and $\sigma _{2}$ but not of $\sigma _{3}$.

The charge-conjugation operation is accomplished by the transformation $\psi
_{c}=\sigma _{1}\psi ^{\ast }$ and the Dirac equation becomes $H_{c}\psi
_{c}=-E\psi _{c}$, with
\begin{equation}
H_{c}=c\sigma _{1}p+\sigma _{3}mc^{2}-\frac{I+\sigma _{3}}{2}\,\Delta -\frac{%
I-\sigma _{3}}{2}\,\Sigma +\sigma _{2}V_{p}\,.  \label{eq5a}
\end{equation}

\noindent One see that the charge-conjugation operation changes the sign of
the energy and of the potentials $V_{t}$ and $V_{p}$. In turn, this means
that $\Sigma$ turns into $-\Delta$ and $\Delta$ into $-\Sigma$. Therefore,
to be invariant under charge conjugation, the Hamiltonian must contain only
a scalar potential.

The chiral operator for a Dirac spinor is the matrix $\gamma^{5}=\sigma
_{1} $. Under the \textit{discrete chiral transformation} the spinor is
transformed as $\psi_{\chi }=\gamma ^{5}\psi $ and the transformed
Hamiltonian $H_{\chi }=\gamma ^{5}H\gamma ^{5}$ reads
\begin{equation}
H_{\chi }=c\sigma _{1}p-\sigma _{3}mc^{2}+\frac{I+\sigma _{3}}{2}\,\Delta +%
\frac{I-\sigma _{3}}{2}\,\Sigma +\sigma _{2}V_{p}.  \label{eq8}
\end{equation}

\noindent This means that the chiral transformation changes the sign of the
mass and of the scalar and pseudoscalar potentials, thus turning $\Sigma$
into $\Delta$ and vice versa. A chiral invariant Hamiltonian needs to have
zero mass and $V_{s}$ and $V_{p}$ zero everywhere.

\subsection{Equations of motion and the Sturm-Liouville problem}

If we now write the spinor $\psi $ in terms of its components $\psi
^{T}=(\psi _{+},\,\psi _{-})$, the Dirac equation gives rise to two coupled
first-order equations for the upper, $\psi _{+}$ and the lower, $\psi _{-}$
components of the spinor:
\begin{equation}
-i\hbar c\psi _{-}^{\prime }+mc^{2}\psi _{+}+\Sigma \psi _{+}-iV_{p}\psi
_{-}=E\psi _{+}  \label{eq11a}
\end{equation}%
\begin{equation}
-i\hbar c\psi _{+}^{\prime }-mc^{2}\psi _{-}+\Delta \psi _{-}+iV_{p}\psi
_{+}=E\psi _{-}\,,  \label{eq11b}
\end{equation}

\noindent where the prime denotes differentiation with respect to $x$. In
terms of $\psi_{+}$ and $\psi_{-}$ the spinor is normalized as $%
\int^{+\infty }_{-\infty }dx(|\psi_{+}|^{2}+|\psi_{-}|^{2})=1$, so that $%
\psi_{+}$ and $\psi_{-}$ are square integrable functions. It is clear from
the pair of coupled first-order differential equations (\ref{eq11a}) and (%
\ref{eq11b}) that $\psi_{+}$ and $\psi_{-}$ have opposite parities if the
Dirac equation is covariant under $x\rightarrow -x$.

For $\Delta =0$ with $E\not=-mc^{2}$, the Dirac equation becomes
\begin{equation}
\psi _{-}=-i\,\frac{\hbar c\psi _{+}^{\prime }-V_{p}\psi _{+}}{E+mc^{2}}\,,
\label{eq15a}
\end{equation}%
\begin{equation}
-\frac{\hbar ^{2}}{2m}\,\psi _{+}^{\prime \prime }\,+\frac{(E+mc^{2})\Sigma
+V_{p}^{2}+\hbar cV_{p}^{\prime }}{2mc^{2}}\psi _{+}=\frac{E^{2}-m^{2}c^{4}}{%
2mc^{2}}\,\psi _{+}\,,  \label{eq15b}
\end{equation}

\noindent and for $\Sigma =0$ with $E\not=mc^{2}$, the Dirac equation
becomes
\begin{equation}
\psi _{+}=-i\,\frac{\hbar c\psi _{-}^{\prime }+V_{p}\psi _{-}}{E-mc^{2}} \label{eq16a}
\end{equation}
\begin{equation}
-\frac{\hbar ^{2}}{2m}\,\psi _{-}^{\prime \prime }\,+\frac{(E-mc^{2})\Delta
+V_{p}^{2}-\hbar cV_{p}^{\prime }}{2mc^{2}}\psi _{-}=\frac{E^{2}-m^{2}c^{4}}{%
2mc^{2}}\,\psi _{-}\,.  \label{eq16b}
\end{equation}

\noindent Either for $\Delta =0$ with $E\not=-mc^{2}$ or $\Sigma =0$ with $%
E\not=mc^{2}$ the solution of the relativistic problem is mapped into a
Sturm-Liouville problem in such a way that the solution can be found by solving
a Schr\"{o}dinger-like problem. In the case of invariance under reflection
through the origin ($x\rightarrow -x$), wave functions with well-defined
parities can be built. Thus, it suffices to study only the
positive half-line and impose adequate boundary conditions on $\psi _{+}$ or $\psi
_{-}$ at the origin and at infinity. Normalizability demands that $\psi _{\pm
}\rightarrow 0$ as $x\rightarrow \infty $. Eigenfunctions and their first
derivatives which are continuous on the whole line with well-defined parities can be
constructed by taking symmetric and antisymmetric linear combinations of $%
\psi _{\pm }$ defined on the positive side of the $x$-axis.

The solutions for $\Delta =0$ with $E=-mc^{2}$ and $\Sigma=0$ with $E=mc^{2}$%
, excluded from the Sturm-Liouville problem, can be obtained directly from
the original first-order equations (\ref{eq11a}) and (\ref{eq11b}). They are

\begin{equation}\label{isolada1}
\begin{array}{c}
\psi_{+} = \psi^{(0)}_{+} \mathrm{exp}\left[ +\int^{x}dy\frac{V_{p}(y)}{%
\hbar c} \right] \\
\\
\psi^{\prime }_{-}+\frac{V_{p}}{\hbar c}\psi_{-} = -\frac{i}{\hbar c}%
(\Sigma+2mc^{2})\psi_{+}%
\end{array}
\end{equation}

\noindent for $\Delta =0$ with $E=-mc^{2}$, and
\begin{equation}\label{isolada2}
\begin{array}{c}
\psi_{-} = \psi^{(0)}_{-} \mathrm{exp}\left[ -\int^{x}dy\frac{V_{p}(y)}{%
\hbar c} \right] \\
\\
\psi^{\prime }_{+}-\frac{V_{p}}{\hbar c}\psi_{+} = -\frac{i}{\hbar c}(\Delta
-2mc^{2})\psi_{-}%
\end{array}
\end{equation}

\noindent for $\Sigma=0$ with $E=mc^{2}$, where $\psi^{(0)}_{+}$ and $%
\psi^{(0)}_{-}$ are normalization constants.

\section{The Coulomb potential}

Let us consider

\begin{equation}\label{eq20}
\Sigma=k_{1}|x|, \quad \Delta=0, \quad V_{p}=k_{2}x.
\end{equation}

As we have seen above, the space inversion does not change $\Sigma $ and $%
\Delta $ but changes $V_{p}$ by $-V_{p}$. The chiral transformation performs
the changes $\Delta \rightarrow \Sigma $, $\Sigma \rightarrow \Delta $, $%
m\rightarrow -m$, and $V_{p}\rightarrow -V_{p}$. Moreover, because $\gamma
^{5}$ interchanges the upper and lower components, the resulting pair of
transformed equations of motion are formally the same, so that their
solutions have the same energy eigenvalues. This symmetry can be clearly
seen from the two equation pairs (\ref{eq15a})-(\ref{eq15b}) and (\ref{eq16a}%
)-(\ref{eq16b}), as well as from the isolated solutions (\ref{isolada1}) and
(\ref{isolada2}), which are converted into each other by this kind of
transformation. This transformation provides a simple mechanism by which one
can go from the results from the $\Sigma=k_{1}|x|,\ \Delta =0,\
V_{p}=k_{2}x$ case to the case when $\Delta =k_{1}|x|$%
, $\Sigma =0$, $V_{p}=k_{2}x$ by just changing
the sign of $m$ in the combinations $(E\pm mc^2)$ and of $k_{2}$ in the relevant expressions.
Notice, however,
that in this latter case the potentials are different and so the eigenenergies will
also be different.

The Dirac spinor corresponding to the isolated solution with $E=-mc^{2}$ is
obtained from Eq. (\ref{isolada1}). Only for $k_{2}>0$ there is a
normalizable Dirac spinor, the upper component vanishes, whereas the lower
component is an even-parity function given by $\psi _{-}=\psi _{-}^{0}%
\mathrm{exp}\left[ -k_{2}x^{2}/(2\hbar c)\right] $. For $E\neq -mc^{2}$, Eq.
(\ref{eq15b}) takes the form

\begin{equation}
-\frac{\hbar ^{2}}{2m}\,\psi _{+}^{\prime \prime }+\left( \frac{1}{2}%
A\,x^{2}+B|x|\right) \psi _{+}=\frac{E^{2}-m^{2}c^{4}-\hbar c\,k_{2}}{2mc^{2}%
}\,\psi _{+}\,,  \label{eq21}
\end{equation}%
where
\begin{equation}
A=\frac{k_{2}^{2}}{mc^{2}},\quad B=\frac{k_{1}}{2mc^{2}}(E+mc^{2}).
\label{A_B}
\end{equation}

As referred above, when $\Delta=k_{1}|x|,
\quad \Sigma=0, \quad V_{p}=k_{2}x$, one obtains the equation
for $\psi_-$ easily from (\ref{eq16b}), so that one has
\begin{equation}
-\frac{\hbar ^{2}}{2m}\,\psi _{-}^{\prime \prime }+\left( \frac{1}{2}%
A'\,x^{2}+B'|x|\right) \psi _{-}=\frac{E^{2}-m^{2}c^{4}+\hbar c\,k_{2}}{2mc^{2}%
}\,\psi _{-}\,,  \label{Sigma=0_geral}
\end{equation}%
where
\begin{equation}
A'=\frac{k_{2}^{2}}{mc^{2}},\quad B'=\frac{k_{1}}{2mc^{2}}(E-mc^{2}).
\label{A'_B'}
\end{equation}

The equations presented are similar to the Dirac equations in 3+1 dimensions,
 with spin and pseudospin symmetries, either with
one-dimensional or radial linear confining potentials, in this last case for s
($\ell=0)$ states.

\subsection{Pure scalar and vector couplings}

In this section we present the solutions of eq.~(\ref{eq21}) when $k_{2}=0$ ($A=0$).
As already mentioned, the solutions of this case and for $\Sigma=0$ were already presented
in \cite{PLA305:100:2002}. Here we review those solutions in a different perspective,
pointing out the relationship between the
$\Delta=0$ and $\Sigma=0$ solutions and also discussing the non-relativistic
limits. This sets a convenient framework for the discussion of the new results presented in the
next subsection.

When $k_{2}=0$ ($A=0$), (\ref{eq21}) reduces to
\begin{equation}
-\frac{\hbar ^{2}}{2m}\,\psi _{+}^{\prime \prime }+B|x|\psi _{+}=\frac{%
E^{2}-m^{2}c^{4}}{2mc^{2}}\,\psi _{+}\,.  \label{casopseu}
\end{equation}

\noindent For this class of potentials, the existence of bound-state
solutions requires $B>0$ and thus, from (\ref{A_B}), one must have
$E\gtrless -mc^{2}\quad \mathrm{for}\quad k_{1}\gtrless 0$.
Eq. (\ref{casopseu}) for $x>0$ turns into the Airy differential equation
\begin{equation}
\frac{d^{2}\psi _{+}}{dz^{2}}-z\psi _{+}=0,  \label{airy}
\end{equation}
where $z=ax+b$, $a=[{k_{1}}\big/({\hbar ^{2}c^{2}})\,(E+mc^{2})]
^{1/3}$, $b=-{a}\,(E-mc^{2})\big/{k_{1}}$,
which has square-integrable solutions expressed in terms of the
Airy functions \cite{ABRAMOWITZ1965}: $\psi_{+}(z)=C\mathrm{Ai}(z)$, where $C$ is a
normalization constant. Continuity of $\psi_{+}$ and its derivative at $x=0$
imply that the homogeneous Dirichlet boundary condition ($\psi \left(
0\right) =0$) must be satisfied for odd-parity solutions whereas the homogeneous
Neumann condition ($\left. d\psi /dx\right\vert _{x=0}=0$) must be satisfied for
even-parity solutions, i.e., $\mathrm{Ai}(b)=0$ for odd-parity and
$\mathrm{Ai}^{\prime }(b)=0$ for even-parity solutions.

These quantization conditions have solutions only for $b<0$ and can be found numerically.
From the definition of $b$ one can see
that $b<0$ always corresponds to $|E|>mc^{2}$ regardless
of the sign of $k_{1}$.
From the roots of $\mathrm{Ai}(b)$ and $\mathrm{Ai}^{\prime }(b)$
we obtain the possible energies as the solutions
of a fourth-degree algebraic equation:
\begin{equation}
(E-mc^2)^3(E+mc^2)-\left( \hbar ck_{1}\right)
^{2}|b|^{3}=E^{4}-2mc^{2}E^{3}+2m^{3}c^{6}E-\left[ m^{4}c^{8}+\left( \hbar ck_{1}\right)
^{2}|b|^{3}\right] =0  \label{eq19b}
\end{equation}
which can be recast in the form (with $E=mc^{2}+mc^2\delta_+ $)
\begin{equation}
\delta_+^{3}(\delta_++2)-D =\delta_+^{4}+2\delta_+^3-D=0
\label{delta_+}
\end{equation}
where $D=[{\hbar k_{1}}\big/({m^2c^3})]^{2}(-b)^{3}=\kappa_{1}^2(-b)^{3}$ and
$\kappa_1=k_1\,{\lambda_c}\big/{(mc^2)^2}=k_1\,\hbar\big/({m^2c^3}($,
$\lambda_c=\hbar/(mc)$ being the reduced Compton wavelength
and $\kappa_1$ the dimensionless strength of the potential.
Since $D$ is positive, there is only one positive solution according to
Descartes\'{} rule of signs for the roots of polynomials
(see,\textit{\ e.g.}, \cite{SALVADORI1952,BEREZIN1965}) and so this is the
only solution for $k_1>0$. For $k_1<0$, since one have $E<mc^2$, we can set
$E=-mc^{2}-mc^2\delta_- $ and find
\begin{equation}
\delta_-(\delta_-+2)^3-D =\delta_-^{4}+6\delta_-^3+12\delta_-^2+8\delta_--D=0 \ .
\label{delta_-}
\end{equation}
Again, there is only a positive root, and therefore the only root.
It is interesting to note
that this result is true whatever the values of the fermion masses
and the coupling constant.
\begin{figure}[!ht]
\begin{center}
\includegraphics[width=8cm]{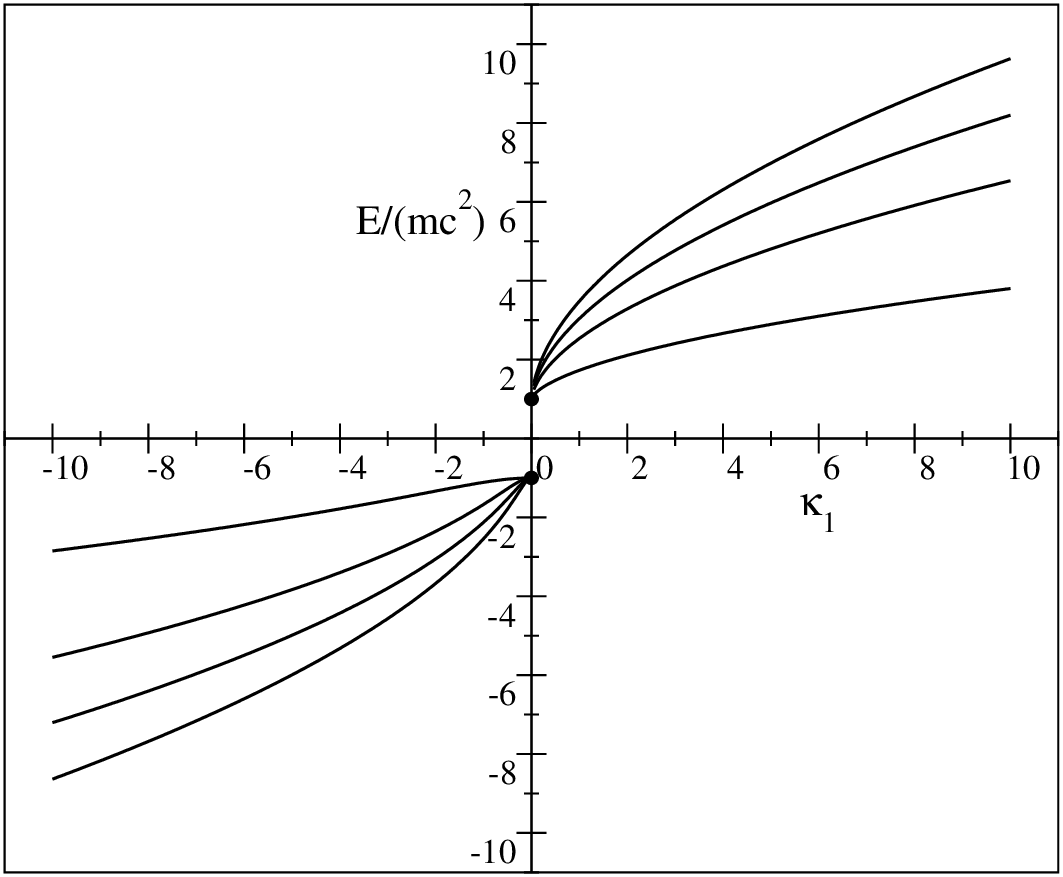}
\caption{First four energy levels as a functions of the dimensionless coupling constant $\kappa_1$
when $k_2=0$ and $\Delta=0$. The black circles represent the
non-bound solutions $E=\pm mc^2$.}
 \label{fig1}
\end{center}
\end{figure}
Fig.~\ref{fig1} shows the behavior of the energies for the four
lowest energy levels as a function of $\kappa_{1}$, when $\Delta=0$.
From Fig.~\ref{fig1} one sees that all
the energies levels for $\kappa_{1}\,(k_1)>0$ emerge from $mc^{2}$ and one finds that
the lowest quantum numbers correspond to the lowest energies, as it should
be for particle energy levels. For $\kappa_{1}\,(k_1)<0$ the spectrum presents a similar
behavior but all the energies levels emerge from $-mc^{2}$ and the highest
energy levels are labeled by the lowest quantum numbers. These energy
levels can be identified with antiparticle levels. This conclusion confirms
what has already been analyzed in \cite{PS75:170:2007,PS77:045007:2008}: the spectrum contains
either particle-energy levels or antiparticle-energy levels depending on the
sign of the coupling constant, but not both kind of levels for a certain
value of coupling constant, as is the case in the 3+1 Coulomb problem
\cite{PRA86:032122:2012}. The same behavior occurs for harmonic oscillator
potentials as reported in \cite{PRC73:054309:2006}.

If one takes the non-relativistic limit by setting $|\kappa_1|\ll 1$, then,
from (\ref{delta_+}), one gets for positive $\kappa_1$ ($\delta_+\ll 1$)
\begin{equation}
2\delta_+^{3}=D \quad\Rightarrow\quad \delta_+=\bigg(\frac D2\bigg)^{1/3} \ ,
\label{delta_+_nr}
\end{equation}
and for negative $\kappa_1$ ($\delta_-\ll 1$), from(\ref{delta_-}),
\begin{equation}
8\delta_-=D \quad\Rightarrow\quad \delta_-=\frac D8 \ .
\label{delta_-_nr}
\end{equation}
In terms of $\kappa_1$, the result for $\delta_-$ (negative energy, anti-fermions) is of higher
order (3 times as much) than $\delta_+$ (positive energy, fermions). From this we may
state that there is no non-relativistic limit for negative energy solutions, since the
solution corresponds to a higher-order term in a $1/(mc^2)$ expansion. Furthermore, one
can check that the solution of the one-dimensional Schr\"odinger equation with a linear
confining potential depends on the strength
of the potential raised to a power of $2/3$,  as is the case for $\delta_+$ in (\ref{delta_+_nr}).

In the case when $\Sigma=0$, i.e., setting $A'=0$ and $k_2=0$ in eq.~(\ref{Sigma=0_geral}),
one gets
\begin{equation}
-\frac{\hbar ^{2}}{2m}\,\psi _{-}^{\prime \prime }+B'|x|\psi _{-}=\frac{%
E^{2}-m^{2}c^{4}}{2mc^{2}}\,\psi _{-}\,\quad{\rm or}\quad
\frac{d^{2}\psi _{-}}{dz'^{2}}-z'\psi _{-}=0\quad{\rm when}\ x>0\,,  \label{casopspin}
\end{equation}
where
$z'=a'x+b'$, $a'=[{k_{1}}\big/({\hbar ^{2}c^{2}})\,(E-mc^{2})]
^{1/3}$, $b'=-{a'}\,(E+mc^{2})\big/{k_{1}}$.
One finds again that one must have $E>mc^2$ or $E<-mc^2$ when $k_1>0$ or $k_1<0$,
respectively. Also one may notice immediately that $a'$ and $b'$ could be obtained from
$a$ and $b$ by changing $m\to -m$, as referred before. So it is not surprising that the solutions of this equation are again Airy functions, the values of $b'$ being its zeros or the zeros of its derivative, depending on to the
parity of $\psi_-$. Therefore, one can set $b'=b$. The eigenvalue
equation, with $k_1>0$, $E=mc^{2}+mc^2\delta'_+ $ is
\begin{equation}
\delta'_+(\delta'_++2)^3-D ={\delta'}_+^{4}+6{\delta'}_+^3+12{\delta'}_+^2+8{\delta'}_+-D=0 \ ,
\label{deltal_+}
\end{equation}
where $D$ is given as before. Eq.~(\ref{deltal_+}) is identical to (\ref{delta_-}) so one can state
that there exists one solution with $E>mc^2$ and that there is no non-relativistic limit for this solution.
For $k_1<0$, i.e., for $E=-mc^2-mc^2\delta'_-<0$, one gets an equation for $\delta'_-$ identical to
(\ref{delta_+}) for $\delta_+$.

\vspace{1mm}
\begin{figure}[!ht]
\begin{center}
\includegraphics[width=8cm]{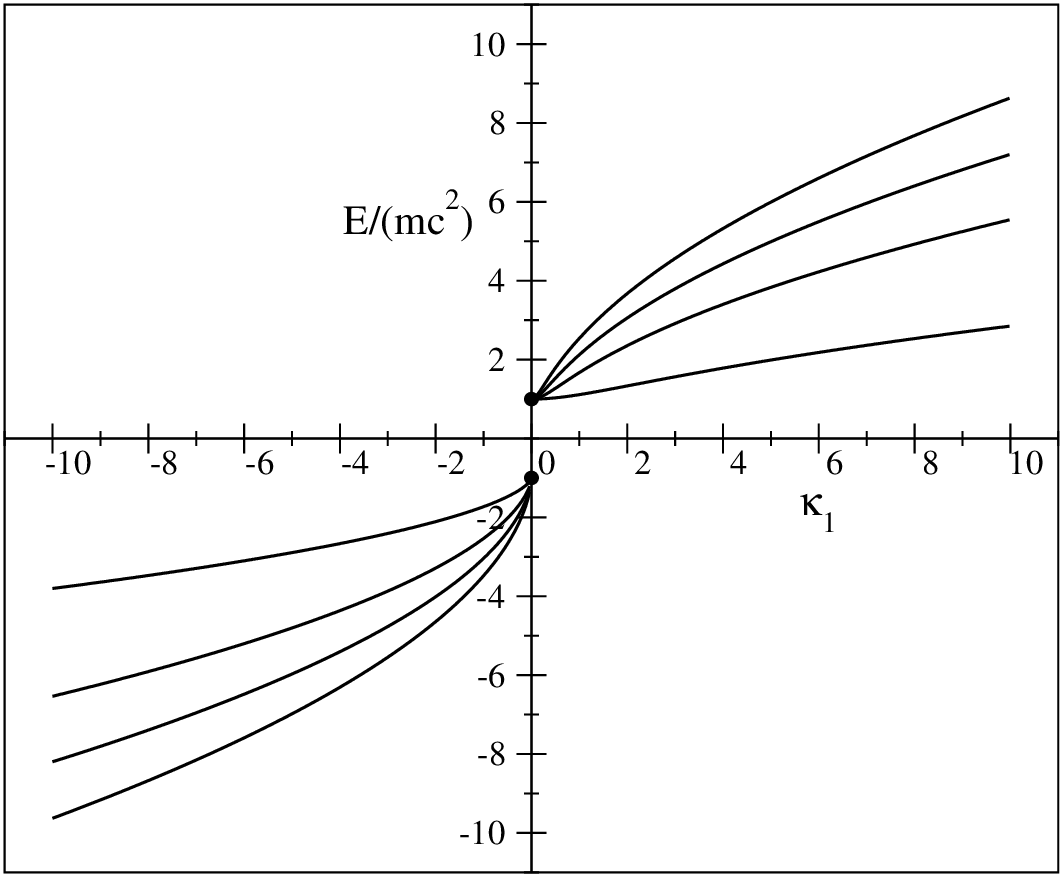}
\caption{First four energy levels as a functions of $\kappa_1$ when $k_2=0$ and $\Sigma=0$.
The black circles represent the non-bound solutions $E=\pm mc^2$.}
 \label{fig2}
\end{center}
\end{figure}

This can be seen from Fig.~\ref{fig2} which shows the behavior of the energies for the four
lowest energy levels as a function of $\kappa_{1}$, when $\Sigma=0$.
Thus we can conclude that the solutions for positive and negative energy when $\Sigma=0$
are reversed with respect to the solutions when $\Delta=0$ in the sense that, for the same value of $|\kappa_1|$,
the energies $|E|$ are the same for positive (negative) $\kappa_1$ for $\Delta=0$ as for negative (positive)
$\kappa_1$ for $\Sigma=0$. Probably the most important result is that one may regard the positive solutions for
$\Sigma=0$ as intrinsically relativistic, since there is no non-relativistic limit. This was already pointed out
in ref. \cite{PRC73:054309:2006} quite generally for any potential $\Delta$ when $\Sigma=0$.
One has also shown that, as in \cite{PRC73:054309:2006} for harmonic oscillator potentials, one is able to find bound solutions
when $\Sigma=0$ for linear confining potentials as is the present case. This agrees with the general finding that in the 3+1 spherically
symmetric case there are only bound solutions for $\Sigma=0$ when the vector and scalar potentials are confining potentials
\cite{PRC87:031301:2013}.

\subsection{The addition of a pseudoscalar coupling}

For $k_{2}\neq 0$ ($A\neq 0$), the equation (\ref{eq21}) ($\Delta=0$) can be cast into
the form
\begin{equation}
-\frac{\hbar ^{2}}{2m}\,\psi _{+}^{\prime \prime }+\frac{A}{2}\left( |x|+%
\frac{B}{A}\right) ^{2}\psi _{+}=\hbar \sqrt{\frac{A}{m}}\left( \nu +\frac{1%
}{2}\right) \,\psi _{+}\,,  \label{eq21a}
\end{equation}

\noindent where%
\begin{eqnarray}
\nonumber
\nu&=&-\frac{1}{2}+\left( \frac{E^{2}-m^{2}c^{4}-\hbar \,ck_{2}}{2mc^{2}}+%
\frac{B^{2}}{2A}\right)\,\frac1\hbar\, \sqrt{\frac{m}{A}}\\
 &=&\frac{E^{2}-m^{2}c^{4}}{2\hbar \,c|k_{2}|}-\frac{1}{2}\left(\frac{k_2}{|k_2|}+1\right)+%
\frac{k_1^2}{|k_2|^3}\,\frac{(E+mc^2)^2}{8\hbar c}\ .  \label{nu}
\end{eqnarray}

The existence of bound solutions of
equation (\ref{eq21a}) is guaranteed by the fact that $A>0$, independently of the sign of $k_2$.
Let us define for positive $x$
\begin{equation}
y=\alpha x+\beta _{\Sigma },\quad \alpha =\sqrt{\frac{2|k_{2}|}{\hbar c}}%
\,,\quad \beta _{\Sigma }=\sqrt{\frac{2|k_{2}|}{\hbar c}}\frac{%
k_{1}(E+mc^{2})}{2k_{2}^{2}},  \label{mudadev}
\end{equation}%
\noindent \noindent so that (\ref{eq21a}) transmutes into
\begin{equation}
\frac{d^{2}\psi _{+}}{dy^{2}}+\left( \nu +\frac{1}{2}-\frac{y^{2}}{4}\right)
\psi _{+}=0\,.  \label{eq2111}
\end{equation}%
The boundary condition $\psi _{+}\rightarrow 0$ as $x\rightarrow \infty $
implies that we must seek solutions of (\ref{eq2111}) which vanish as $%
y\rightarrow \infty $. The particular solution of (\ref{eq2111}) which
vanishes for very large positive values of $y$ is called a parabolic
cylinder function, it is denoted by $D_{\nu }(y)$. Now, the boundary
conditions at $x=0$ lead to the quantization conditions
\begin{eqnarray}
D_{\nu }\left( \beta _{\Sigma }\right)  &=&0\quad \text{\textrm{%
for\thinspace \thinspace odd-parity\thinspace \thinspace solutions}}  \notag
\\[-2mm]
&&  \label{cond_front_D_nu} \\[-2mm]
D_{\nu }^{\prime }\left( \beta_{\Sigma }\right)  &=&0\quad \text{\textrm{%
for\thinspace \thinspace even-parity\thinspace \thinspace solutions.}}
\notag
\end{eqnarray}
These equations are non-explicit equations for the energy $E$, since the dependence on $E$
comes about not only through $\beta_{\Sigma}$ but also through $\nu$. From the properties of the real zeros
of the the parabolic cylinder functions \cite{FRANK2010} one must have $\nu>0$ for $D_{\nu }$
to have real zeros, and their derivatives $D_{\nu }^{\prime }$ have real zeros for $\nu\gtrsim -0.21$, as checked with
the program MATHEMATICA.
If $\nu=0$ $D_{\nu }^{\prime }$ has a zero at the origin.
This means that the following condition for the energies and strengths of the potentials must be
satisfied
\begin{equation}
\label{nu_condition}
\nu =\frac{E^{2}-m^{2}c^{4}}{2\hbar \,c|k_{2}|}-\frac{1}{2}\left(\frac{k_2}{|k_2|}+1\right)+%
\frac{k_1^2}{|k_2|^3}\,\frac{(E+mc^2)^2}{8\hbar c}\gtrsim -0.21 \ .
\end{equation}
When $\nu\geq 0$, solutions with both parities exist. Also, depending on the particular value of $\nu$, there are both positive and
negative zeros of $D_{\nu }$ and $D_{\nu }^{\prime }$, so that, from (\ref{mudadev}), there can exist
both positive and negative energy solutions for a particular pair of $(k_1,k_2)$ values. As mentioned before, in a purely confining
scalar and/or vector potential this is not possible.

The fact that the zeros of $D_{\nu }$ \cite{FRANK2010} and also $D_{\nu }^{\prime }$ are bounded by the value of $\nu$
further restricts the existence of solutions in this case. This is because the values of $\beta_{\Sigma}$ and $\nu$,
for fixed $\kappa_1$ and $\kappa_2$, are themselves connected by their expressions in eqs. (\ref{mudadev}) and  (\ref{nu_condition}) by elimination of the energy $E$.
Numerical tests show that the ratio $\kappa_1/\kappa_2$ cannot be big in order to have solutions for this problem.

The special case of $k_{1}=0$ is worth mentioning. In this case,
$\beta_{\Sigma }=0$ and $\nu\equiv n=0,1,2,\ldots$ is an integer.
We have then $D_{{n}}(y)=2^{{-n/2}}e^{{-\frac{1}{4}y^{2}}}\mathop{H_{{n}}\/}%
\nolimits\!\left(y/\sqrt{2}\right)$, where $H_n$ are the Hermite polynomials of degree $n$
\cite{FRANK2010}. This corresponds to the pseudoscalar Coulomb
potential, already analyzed in \cite{PLA318:40:2003} or, viewed in another way, the so-called
Dirac oscillator in 1+1 dimensions studied in \cite{PRC73:054309:2006}. The spectrum has both positive and negative energies
for a particular value of $k_2$. Defining $\kappa_2=k_2/(\hbar{m^2c^3})$ as the dimensionless strength of the pseudoscalar
coupling, one has explicitly
\begin{equation}
\label{ener_DO}
{E_n}^2=m^2c^4\big[1+\kappa_2+|\kappa_2|(2n+1)\big] \ .
\end{equation}

\begin{figure}[!ht]
 \begin{center}
\includegraphics[width=8cm]{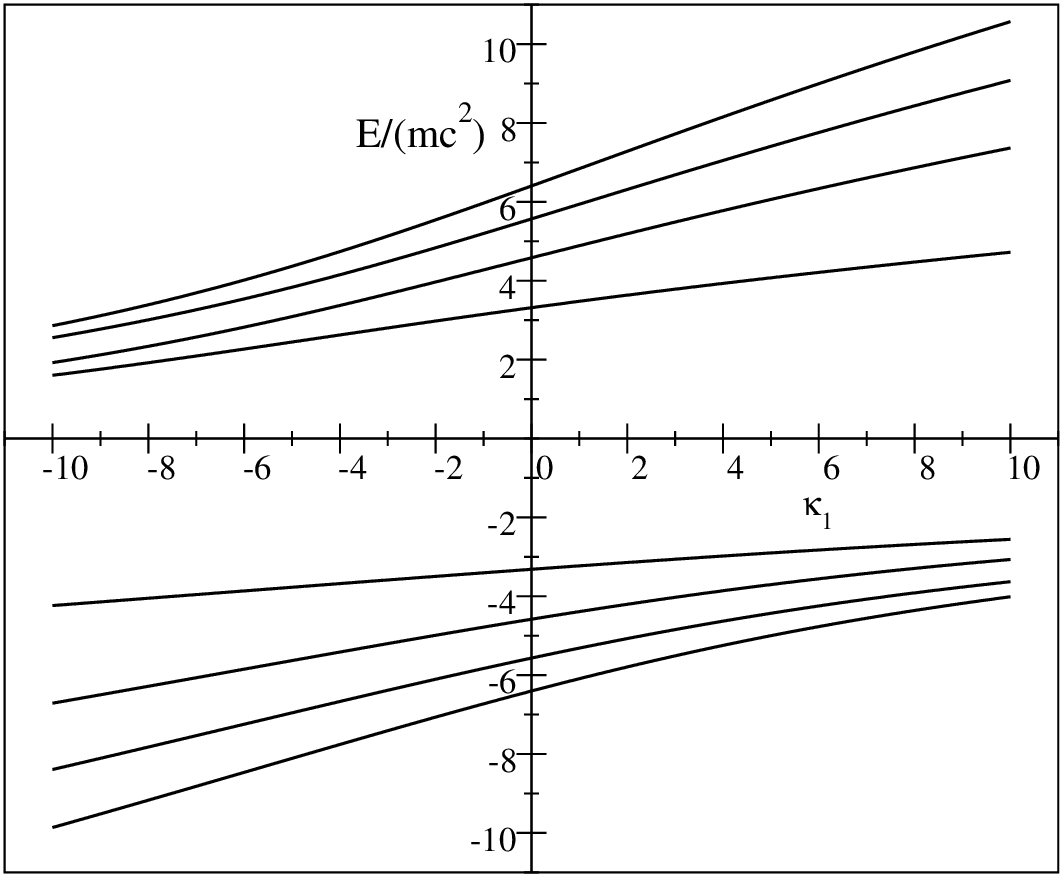}
\caption{First four energy levels as a functions of $\kappa_1$ when $\kappa_2=5$ and $\Delta=0$.}
 \label{fig3}
 \end{center}
\end{figure}

\begin{figure}[!ht]
\begin{center}
\includegraphics[width=8cm]{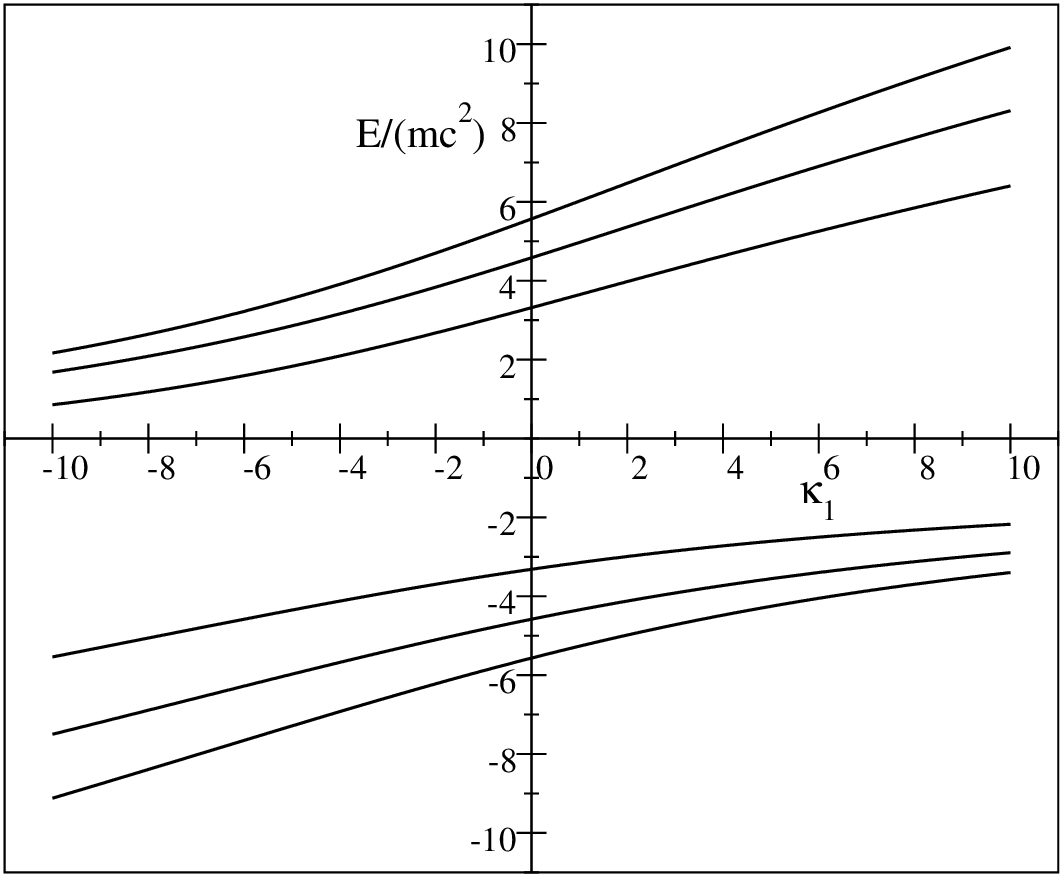}
\caption{First three energy levels as a functions of $\kappa_1$ when $\kappa_2=-5$ and $\Delta=0$.}
 \label{fig4}
\end{center}
\end{figure}

We used this solution as a starting point for finding solutions for this problem, fixing $\kappa_2$ and allowing $\kappa_1$ to vary.
We chose a value of $\kappa_2$ large enough in order to keep the ratio $\kappa_1/\kappa_2$ sufficiently small for the range of $\kappa_1$ values considered, since, as referred before, otherwise one cannot get solutions to this problem. Because of this restriction, the complementary study (fixing $\kappa_1$ and allowing $\kappa_2$ to vary, starting from the $\kappa_2=0$ solution) is not feasible.
Note that the quantum number $n$ in (\ref{ener_DO}) defines the parity of the solutions of the Dirac oscillator and, by continuity,
the parity of the solutions in the present case. For even or odd $n$ one has even or odd solutions, respectively. Furthermore, for
 $\kappa_2>0$, $n$ takes the values $0,1,\ldots,$ but for $\kappa_2<0$ the value $n=0$ is excluded \cite{PRC73:054309:2006}.  In Figs.~\ref{fig3} and \ref{fig4} we plot the solutions found numerically for $\kappa_2=5$ and Dirac oscillator quantum number $n=0,1,2,3$ and for $\kappa_2=-5$ and $n=1,2,3$, respectively.

Now, let us present, as an example of the rule stated at the beginning of
the section, the quantization conditions for $\Delta =k_{1}|x|$, $\Sigma =0$%
, $E\neq mc^{2}$ and $V_{p}=k_{2}x$. These are
\begin{eqnarray}
D_{\mu }\left( \beta _{\Delta }\right) &=&0\quad \text{\textrm{for\thinspace
\thinspace odd-parity\thinspace \thinspace solutions}}  \notag \\[-2mm]
&&  \label{eq26b} \\[-2mm]
D_{\mu }^{\prime }\left( \beta _{\Delta }\right) &=&0\quad \text{\textrm{%
for\thinspace \thinspace even-parity\thinspace \thinspace solutions,}}
\notag
\end{eqnarray}

\noindent where%
\begin{equation}
\beta _{\Delta }=\sqrt{\frac{2|k_{2}|}{\hbar c}}\frac{k_{1}(E-mc^{2})}{%
2k_{2}^{2}} \ ,
\end{equation}%
and $\mu$ is given by
\begin{equation}
\label{mu}
\mu =\frac{E^{2}-m^{2}c^{4}}{2\hbar \,c|k_{2}|}+\frac{1}{2}\left(\frac{k_2}{|k_2|}-1\right)+%
\frac{k_1^2}{|k_2|^3}\,\frac{(E-mc^2)^2}{8\hbar c} \ .
\end{equation}

The second-order equation
to solve in this case ($\Sigma =0$) is the one for the lower component, Eq. (\ref%
{eq16b}), because the chiral transformation interchanges the upper and lower
components. The upper component can be obtained from the corresponding
first-order equation Eq. (\ref{eq16a}). Note that for massless particles,
the cases $\Delta =0$ and $\Sigma =0$ have the same spectrum with the sign
of $k_{2}$ reversed.

\begin{figure}[!ht]
\begin{center}
\includegraphics[width=8cm]{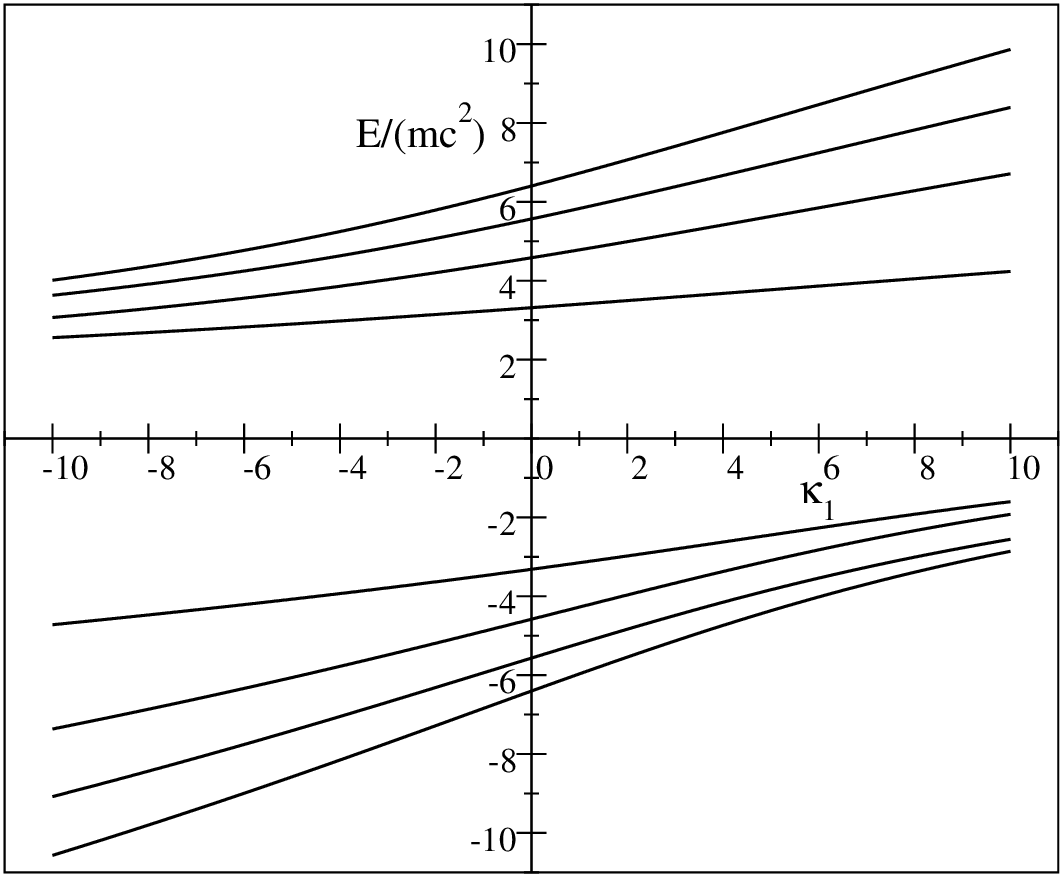}
\caption{First four energy levels as a functions of $\kappa_1$ when $\kappa_2=-5$ and $\Sigma=0$.}
 \label{fig5}
 \end{center}
\end{figure}
In Figure \ref{fig5} we show the solutions for $\Sigma=0$ found numerically for $\kappa_2=-5$ and Dirac oscillator quantum number $n=0,1,2,3$ (note that, when solving for $\Sigma=0$ and taking the limit $\kappa_1\to 0$ in the expression for the energy of Dirac oscillator
(\ref{ener_DO}), the sign of $\kappa_2$ is reversed).

Comparing with Figure \ref{fig3}, we see that the plots are identical if we reverse
both the vertical and horizontal axes, i.e., we get the same solutions as in the case of $\kappa_2=5$ and $\Delta=0$
if we reverse the sign of the energy  and of $\kappa_1$, respectively. This is because $\beta_\Sigma$ turns into $\beta_\Delta$ if we make the changes $k_1\to -k_1$ and $E\to -E$ and, on the other hand, $\nu$ turns into $\mu$  when $k_2\to -k_2$ and $E\to -E$, while it is left unchanged by the change of the sign of $k_1$.
This illustrates the fact that the case $\Sigma =0$ can also be obtained from the $\Delta =0$ case by applying
the charge-conjugation transformation. We recall that this transformation
performs the changes $E\rightarrow -E$, $\Delta \rightarrow -\Sigma $ $%
V_{p}\rightarrow -V_{p}$, i.e., the changes $k_{1}\rightarrow -k_{1}$ and $%
k_{2}\rightarrow -k_{2}$.

Thus we see that, by adding the pseudoscalar coupling, one allowed for
the possibility of having bounded states with both positive- and
negative-energy solutions in a system with either $\Delta=0$ or $\Sigma=0$, which can be relevant
in strong interaction physics where these linear confining potentials can appear.

\section{Conclusions}

In this paper we have computed and described in detail the bound-state
solutions of the 1+1 Dirac equation with a Coulomb potential with the most
general Lorentz structure in spin and pseudospin symmetry conditions.
We found that there exist bounded solutions for
both particles and antiparticles but not for all the range of the coupling strengths.

At the same time, this work illustrates some general conclusions drawn in previous works
about spin and pseudospin symmetry, namely that one can obtain the solutions for $\Sigma =0$
 from the $\Delta =0$ case, using the
charge-conjugation and chiral transformations.

Another important conclusion is that, as previously shown for harmonic oscillator potentials,
one is able to get bound states in pseudospin symmetry conditions ($\Sigma =0$, with or without
a pseudoscalar potential), which seems
to confirm the assertion, made in the context of the 3+1 Dirac equation with radial potentials,
that the asymptotic behavior of the potentials (confining in this case) is a crucial
feature that allows to have such bound states.

To finish with, one should also refer that these findings may be applied to the one-dimensional
linear potential problem in a 3+1 Dirac equation and in principle in the spherically
symmetric problem also in 3+1 dimensions, provided one has $\ell=0$ or $\tilde\ell=0$ in the
spin symmetric ($\Delta=0$) or the pseudospin symmetric ($\Sigma=0$) cases, respectively,
$\ell$ and $\tilde\ell$ being the orbital and pseudospin orbital angular momentum quantum numbers,
respectively \cite{PRC69:034318:2004}. As such, these results might be of relevance to quarkonium phenomenology in 3+1
dimensions.

\section*{Acknowledgments}
This work was supported in part by means of funds provided by CAPES and CNPq.
P. Alberto would like also to thank the Universidade Estadual
Paulista, Guaratinguet\'a Campus, for supporting his stays in its
Physics and Chemistry Department. L. B. Castro thanks CNPq (grants 455719/2014-4 and 304105/2014-7) for partial support.

\section*{References}

\bibliography{mybibfile_stars2}

\end{document}